\documentclass[aps,prd,twocolumn,groupedaddress,showpacs,showkeys,draft]{revtex4-1}
\usepackage{graphicx}% Include figure files
\usepackage{dcolumn}% Align table columns on decimal point
\usepackage{bm}% bold math
\usepackage{amssymb,amsmath,latexsym,amsfonts}
\begin{document}

%\preprint{APS/123-QED}

\title{Landua Criterion in a Bose--Condensed Sodium Gas}% Force line breaks with \\

 \author{A. Camacho}
 \email{acq@xanum.uam.mx} \affiliation{Departamento de F\'{\i}sica,
 Universidad Aut\'onoma Metropolitana--Iztapalapa\\
 Apartado Postal 55--534, C.P. 09340, M\'exico, D.F., M\'exico.}


\date{\today}% It is always \today, today,

\begin{abstract}
In light of the experimental evidence for the existence of a
superfluidity region in a Bose--condensed sodium gas a theoretical
model is put forward. It will be shown that the predictions of the
present work do match with the extant measurement readouts. As a
byproduct we also calculate the speed of sound and compare it
against the current experimental results.

\end{abstract}

\maketitle

%****************************************************************************
The first experimental evidence concerning the phenomenon of
superfluidity can be tracked down to the work of Kamerling Onnes in
the year of 1911 with helium when it was found that if cooled below
$2.2 ~^{\circ}K$ $He$ did not contract but rather expand
\cite{Onnes}. Since then the amount of theoretical and experimental
work has been able to provide a coherent picture to the subjacent
Physics \cite{Khalatnikov, Nozieres, Griffin}. Another phenomenon,
which also emerged in the last century, is the Bose--Einstein
Condensation (BEC) the one, as in the case of superfluidity, is
inherently related to the presence of very low temperatures. The
need for low temperatures for the appearance of these two effects
leads us to the obvious question concerning a possible relation
between them. It was Fritz London \cite{London} who put forward the
idea of a connection between these two effects stating that the
transition from $He~I$ (the high temperature phase of liquid helium)
and $He~II$ (the low temperature phase) should be considered an
example of a BEC. Even more, he suggested that in $He~II$ a
macroscopic quantum current of matter could be present, i.e., he
introduced the idea that BEC and superfluidity could appear,
simultaneously, in a system.

It has to be stressed that this understanding cannot be considered a
closed issue \cite{Yukalov}, and the answer to this aforementioned
interrogant is that, though there is a close relationship, it is not
a unique one. Indeed, we may state that BEC is neither necessary nor
sufficient for the presence of superfluidity. For instance, and
ideal Bose--Einstein condensate shows no superfluidity, and, as a
counterpart a two--dimensional superfluid cannot condensate
\cite{Ueda}.

Landau (within the two--fluid model proposed by Tisza \cite{Tisza})
introduced the concept of elementary excitations \cite{Landau1} as a
fundamental element in the description of the behavior of $He~II$.
Landau argued \cite{Landau2} that the normal fluid (the
non--superfluid component) could be regarded as a dilute gas whose
components are weakly--interacting elementary excitations which move
in a background defined by the superfluid component. Accordingly,
the phenomenon of superfluidity appears if the velocity of the
corresponding flow lies below a certain threshold value given by

\begin{equation}
v_{(crit)}= \min\Bigl(\frac{\epsilon(p)}{p}\Bigr). \label{Landau}
\end{equation}

Here $\epsilon(p)$ denotes the energy of an elementary excitation
and $p$ its corresponding momentum. If the velocity is larger, then
the microscopic rugosities of the walls of the container will
scatter the particles of the fluid entailing the loss of kinetic
energy of the fluid, i.e., viscosity appears.

In the experimental realm the quest for this critical velocity has
been carried out in a sodium--BEC, and the results show a possible
velocity threshold located around the value of $1.6mm/s$
\cite{Raman}. The use of sodium--condensed gas in the experimental
context is related not only to the aforementioned case but also to
the excitation of phonons by light scattering \cite{Stamper} and the
propagation of sound \cite{Andrews, Andrews1}.

The main purpose of the present work is to obtain a prediction for
the critical velocity for a BEC. The deduced model will be compared
against the reported measurement readouts \cite{Raman}, and, in
addition, the speed of sound for a sodium--condensed gas will be
found and compared with the current experimental results
\cite{Andrews, Andrews1}.

From a fundamental point of view our mathematical model can be
defined by an $N$--particle Hamiltonian the one in the formalism of
second quantization is \cite{Ueda}

\begin{eqnarray}
\hat{H}= \int d\vec{r}\Bigl[-\hat{\psi}^{\dagger}(\vec{r}, t)
\frac{\hbar^2}{2m}\nabla^2\hat{\psi}(\vec{r}, t)\nonumber\\
+ V(\vec{r})\hat{\psi}^{\dagger}(\vec{r}, t)\hat{\psi}(\vec{r}, t)\nonumber\\
+\frac{U_0}{2}\hat{\psi}^{\dagger}(\vec{r},
t)\hat{\psi}^{\dagger}(\vec{r}, t) \hat{\psi}(\vec{r},
t)\hat{\psi}(\vec{r}, t)\Bigr]. \label{Hamilton1}
\end{eqnarray}

In this Hamiltonian $\hat{\psi}^{\dagger}(\vec{r}, t)$ and
$\hat{\psi}(\vec{r}, t)$ represent
 bosonic creation and annihilation operators, respectively. It is restricted to low energies
and momenta and implies, as a consequence of the aforementioned
conditions, that the interaction among the particles is, as usual,
codified by the scattering length parameter $a$, i.e.,
$U_0=\frac{4\pi a\hbar^2}{m}$. The trapping potential $V(\vec{r})$,
for our case, corresponds to an isotropic harmonic oscillator whose
frequency reads $\omega$. In addition, there are $N$ particles in
the gas, each of them with mass $m$, the volume occupied by the
system is $V$.

Our mathematical assumptions are:

(i) Only the ground and the first excited states are populated. This
condition can be justified recalling that for a bosonic system, with
chemical potential $\mu$ and energy levels of single--particle
$\epsilon$, the occupation number in thermal equilibrium is given by
\cite{Pethick} ($\beta=1/(\kappa T$))

\begin{eqnarray}
<n_{(\epsilon)}>=\frac{1}{e^{(\epsilon-\mu)\beta}-1}.\label{Occunumber}
\end{eqnarray}

It is readily seen that we deal with a monotonic decreasing function
of $\epsilon$, and this feature justifies the present assumption.

(ii) The mathematical description of the two occupied states will be
done resorting to the Hartree approximation, in which the ground
state of the interacting system is deduced by a
Ginzburg--Pitaevski--Gross energy functional \cite{Gross1}, and it
entails that the ground state wavefunction corresponds to the case
of a harmonic oscillator situation but the frequency is modified due
to the fact that the system has a non--vanishing scattering length
\cite{Baym}, such that the fundamental length parameter reads.

\begin{equation}
R= \Bigl(\frac{2}{\pi}\Bigr)^{1/10}\Bigl(\frac{Na}{l}\Bigr)^{4/5}l.
\label{Radius1}
\end{equation}

Here $l$ is the radius related to the trap given by the isotropic
harmonic oscillator, namely,

\begin{equation}
l= \sqrt{\frac{\hbar}{m\omega}}. \label{Radius2}
\end{equation}

Clearly this condition implies an effective frequency

\begin{equation}
\tilde{\omega}= \frac{\hbar}{mR^2}. \label{Freq1}
\end{equation}

Usually the experimental conditions entail $R>l$ \cite{Raman} and,
in consequence, $\tilde{\omega}<\omega$.

In other words, the order parameter related to the particles in the
ground state is provided by

\begin{eqnarray}
\psi_{(0)}(\vec{r})=\sqrt{\frac{N_{(0)}}
{(R\sqrt{\pi})^3}}\exp{\Bigl[-\frac{r^2}{2R^2}\Bigr]}.\label{Groustate}
\end{eqnarray}

In this last expression $N_{(0)}$ denotes the number of particles in
the lowest energy state. The presence of a non--vanishing scattering
length entails that in the ground state not all the particles can
have zero--momentum, the reason for this lies in the fact that the
two--body interaction mixes in components with atoms in other states
\cite{Pethick} and

\begin{eqnarray}
N_{(0)}=N\Bigl[1-\frac{8}{3}\sqrt{\frac{Na}{\pi
V}}\Bigr].\label{Depletion}
\end{eqnarray}

Clearly,

\begin{eqnarray}
N_{(0)}= \int\bigl(\psi_{(0)}(\vec{r})\bigr)^2d^3r,\label{Norm1}
\end{eqnarray}

\begin{eqnarray}
V=\frac{4\pi}{3}R^3.\label{Volume1}
\end{eqnarray}

 The wavefunction of the first excited state will be
considered as the first excited state of an isotropic oscillator
related to frequency given by (\ref{Freq1}) and, due to our
symmetry, it has three possibilities, all with the same mathematical
structure, namely,

\begin{eqnarray}
\psi^{(i)}_{(1)}(\vec{r})=\frac{8}{\sqrt{27\pi}}\sqrt{\frac{N}{V}\sqrt{\frac{Na}{\pi
V}}}\frac{x^{(i)}}{R}\exp{\Bigl[-\frac{r^2}{2R^2}\Bigr]}.\label{Excstate}
\end{eqnarray}

Here $x^{(1)}=x$, $x^{(2)}=y$, and $x^{(3)}=z$.

Of course, (\ref{Excstate}) must be related to the total number of
particles in excited states ($N_{(e)}=\frac{8}{3}\sqrt{\frac{Na}{\pi
V}}$), a condition that becomes \cite{Pethick}

\begin{eqnarray}
N_{(e)}=\int\Bigl[\sum_{i=1}^{3}
\bigl(\psi^{(i)}_{(1)}(\vec{r})\bigr)^2\Bigr]d^3r.\label{Norm2}
\end{eqnarray}

Having stated our assumptions we proceed to compute the speed of
sound and the critical velocity. The energy of the ground state in
its three possibilities, i.e., kinetic energy, due to the trap, and
interaction are \cite{Baym, Pethick}

\begin{eqnarray}
\frac{\hbar^2}{2m}\int\Bigl(\nabla\psi_{(0)}(\vec{r})\Bigr)^2d^3r=\frac{3\hbar^2}{4mR^2}N_{(0)}.\label{Kinen}
\end{eqnarray}

\begin{eqnarray}
\int V(\vec{r})\psi_{(0)}(\vec{r})d^3r=
\frac{3}{4}m\omega^2R^2N_{(0)}.\label{Trapen}
\end{eqnarray}

\begin{eqnarray}
\frac{U_0}{2}\int\Bigl(\psi_{(0)}(\vec{r})\Bigr)^4d^3r=
\frac{U_0}{\sqrt{32\pi^3}R^3}N^2_{(0)}.\label{Inten}
\end{eqnarray}

The energy of the ground state (here denoted by $E_{(0)})$, no
elementary excitations are present, is the sum of the last three
expressions. The corresponding pressure ($P_{(0)}=-\partial
E_{(0)}/\partial V$) is

\begin{eqnarray}
P_{(0)}=\frac{4\pi\hbar^2
N}{mV^{5/3}}\Bigl\{\frac{1}{8\pi}\bigl(\frac{4\pi}{3}\bigr)^{2/3}\Bigl[1+\frac{3}{4}N_{(e)}\Bigr]\nonumber\\
+\Bigl[\frac{1}{\sqrt{18\pi}}\bigl(1-2N_{(e)}\bigr)\bigl(1-N_{(e)}\bigr)\nonumber\\
+\frac{\sqrt{2\pi}}{4\pi}\bigl(1-\frac{3}{2}N_{(e)}\bigr)
\Bigr]\frac{Na}{V^{1/3}} \Bigr\}.\label{Pressure0}
\end{eqnarray}

If $v_{(0)}$ denotes the speed of sound related to the last
expression ($v^2_{(0)}=-(V^2/mN)\partial P_{(0)}/\partial V$) we are
led to

\begin{eqnarray}
v^2_{(0)}=\frac{\hbar^2}{m^2}\frac{4\pi
Na}{V}\Bigl\{\sqrt{\frac{2}{9\pi}}\Bigl[1-\frac{15}{4}N_{(e)}
+4N^2_{(e)} \Bigr]\nonumber\\
+\frac{1}{\sqrt{2\pi}}\Bigl[1-\frac{15}{8}N_{(e)}\Bigr] +
\frac{5}{12\pi}\bigl(\frac{4\pi}{3}\bigr)^{2/3}\frac{V^{1/3}}{Na}\bigl(1\nonumber\\
+\frac{39}{40}N_{(e)}\bigr)\Bigr\}.\label{Sound0}
\end{eqnarray}

For the case of sodium \cite{Andrews, Andrews1} the physical
parameters are: $m=36.8\times 10^{-27}kg$, $a=2.75\times 10^{-9}m$,
$N=5\times 10^6$. Considering one of the reported peak densities of
the condensate, namely, $1\times 10^{14}cm^{-3}$, we obtain, in the
roughest approximation from our calculations, $6.59~mm/s$. Clearly,
our prediction is in good agreement with the experiment
\cite{Andrews}.

We now proceed to compute the lowest energy and momentum of the
elementary excitations, physical parameters required for the
deduction of the critical velocity \cite{Nozieres}. The deduction of
the energy of an elementary excitation and of its corresponding
momentum requires the knowledge of the energy of a single--particle
in the first excited state \cite{Ueda}. Our assumptions entail that
the thermal cloud contains particles subject to an isotropic
harmonic oscillator whose frequency is (\ref{Freq1}) therefore the
energy of an excited particle is given by this assumption and easily
calculated as a function of the effective frequency of our
variational procedure

\begin{eqnarray}
\tilde{\epsilon}= \frac{5}{2}\hbar\tilde{\omega}.\label{excene1}
\end{eqnarray}

According to Bogoliubov \cite{Ueda, Bogoliubov} the energy of an
elementary excitation, here denoted by  $\epsilon$, is a function of
the energy of the excited particles of the BEC, namely,

\begin{eqnarray}
\epsilon=\sum\sqrt{(\tilde{\epsilon})^2+\frac{2NU_{(0)}}{V}\tilde{\epsilon}}.\label{excene2}
\end{eqnarray}

The energy of all the elementary excitations turns out to be
\cite{Ueda, Bogoliubov}

\begin{eqnarray}
\tilde{E}=\sum\sqrt{(\tilde{\epsilon})^2+\frac{2NU_{(0)}}{V}\tilde{\epsilon}}<\tilde{n}_{\epsilon}>.\label{Excene2}
\end{eqnarray}

Here $<\tilde{n}_{\epsilon}>$ denotes the occupation number of the
elementary excitations with energy ${\epsilon}$. The relation
between the occupation numbers of particles and elementary
excitations is \cite{Ueda}

\begin{eqnarray}
<\tilde{n}_{\epsilon}>=
\frac{<n_{\epsilon}>}{1+<n_{\epsilon}>}.\label{Occunumber3}
\end{eqnarray}

At this point, for the sake of simplicity, we resort to the
experimental values related to the detection of a critical velocity
in a sodium condensed gas \cite{Raman} in which the occupation
number of the particles in the first excited state fulfill the
condition $N_{(e)}\sim 10^2>1$, and, in consequence,
$<\tilde{n}_{\epsilon_{(1)}}>=1$. Our assumptions imply
$<\tilde{n}_{\epsilon_{(i)}}>=0, ~~\forall i>1$. Indeed, we have
considered that the thermal cloud is comprised by particles which
occupy only the first excited state, in other words,
$<n_{\epsilon_{(i)}}>=0,~~\forall i>1$. Introducing this condition
into (\ref{Occunumber3}) leads us to the aforementioned result for
the occupation number of the elementary excitations.

Casting (\ref{excene2}) in terms of the effective volume $V=4\pi
R^3/3$, and using (\ref{Radius1}),  (\ref{Radius2}), and
(\ref{Freq1}), we have that the energy of our elementary excitation
is

\begin{eqnarray}
\epsilon=\Bigl(\frac{4\pi}{3}\Bigr)^{1/3}\frac{\hbar^2}{mV^{2/3}}\sqrt{\frac{25}{4}\Bigl(\frac{4\pi}{3}\Bigr)^{2/3}+
\frac{20\pi Na}{V^{1/3}}}.\label{excene3}
\end{eqnarray}

We must now find the momentum of this elementary excitation.
Elementary excitations, which define the normal component of the
fluid fluid, can be regarded as a bosonic  gas whose components are
weakly--interacting and moving in a region in which a constant
potential exists, and this potential is defined by a mean field
approach \cite{Ueda}. According to this interpretation we may
rewrite (\ref{excene3}) in the same form as in the case in which our
BEC is a homogeneous one \cite{Ueda}. In other words, we cast our
last expression in the following form

\begin{eqnarray}
\epsilon=\frac{\hbar^2k}{2m}\sqrt{k^2+ \frac{16\pi
Na}{V}}.\label{excene4}
\end{eqnarray}

Clearly, (\ref{excene4}) allows us to deduce the wavenumber related
to our elementary excitation and, in consequence, its momentum.
Indeed, we have for these two physical variables, respectively, that

\begin{eqnarray}
k=\Bigl(\frac{4\pi}{3}\Bigr)^{1/3}\sqrt{5}\frac{1}{V^{1/3}}
,\label{elemwavenum}
\end{eqnarray}

\begin{eqnarray}
p=\Bigl(\frac{4\pi}{3}\Bigr)^{1/3}\sqrt{5}\frac{\hbar}{V^{1/3}}
.\label{elemmom}
\end{eqnarray}

Resorting to Landau criterion (\ref{Landau}) we obtain that the
critical velocity is given by

\begin{eqnarray}
v_{(crit)}=\frac{1}{\sqrt{5}}\frac{\hbar}{mV^{1/3}}\sqrt{\frac{25}{4}\Bigl(\frac{4\pi}{3}\Bigr)^{2/3}+
\frac{20\pi Na}{V^{1/3}}}.\label{crit2}
\end{eqnarray}

The experimental parameters \cite{Raman} are a critical speed of
$v^{(e)}_{(crit)}= 1.6~mm/s$. In addition, the number of particles
in this experiment has a minimum of $N=3\times 10^6$ and a maximum
of $N=12\times 10^6$, and for the evaluation of our expression we
will take the arithmetic average, i.e., $N=7.5\times 10^6$. The
effective volume is that of an ellipsoid whose axes are
$l_1=45\times 10^{-6}m$ and $l_1=150\times 10^{-6}m$ such that
$V=\frac{4\pi}{3}l^2_1l_2$.

Introducing these values into (\ref{crit2}) entails

\begin{eqnarray}
v^{(m)}=1.95~mm/s.\label{crit3}
\end{eqnarray}

The reported critical speed is \cite{Raman}

\begin{eqnarray}
v^{(e)}= 1.6~mm/s
\end{eqnarray}

The ensuing error is less that 18 percent
\begin{eqnarray}
\vert v^{(e)}-v^{(m)}\vert/\Bigl(v^{(m)}\Bigr)=0.179.\label{error1}
\end{eqnarray}

Since the number of particles in the corresponding experiment varies
from $N=3\times 10^6$ to $N=12\times 10^6$ \cite{Raman} the
associated values for the critical speed go from $1.24~mm/s$ to
$2.48~mm/s$. If $N=5\times 10^6$, then $v^{(m)}= v^{(e)}$.

In conclusion, we have put forward a theoretical model for the
deduction of the critical velocity in a sodium--condensed gas. This
threshold speed has been computed and compared against the extant
experimental results, having a good agreement between them. A more
precise evaluation of the present idea requires a better knowledge
of the value of $N$ employed in the experiment. Previous works offer
larger critical velocities which have a bigger error than the one
here deduced \cite{Raman, Frisch}, when compared to the experimental
result.

\end{document}